\documentclass[twocolumn,showpacs,prb]{revtex4}
\usepackage{hyperref}

\usepackage[utf8x]{inputenc}

\usepackage{units}
\usepackage{amsthm}
\usepackage{amsmath}
\usepackage{graphicx}
\usepackage{setspace}
\usepackage{amssymb}

\begin{document}

\title{Attosecond nanoplasmonic streaking of localized fields near metal
nanospheres}
\author{Frederik S\"{u}{\ss}mann and Matthias F. Kling}
\affiliation{Max-Planck-Institut f\"{u}r Quantenoptik, Hans-Kopfermann-Straße 1,
D-85748 Garching, Germany}

\begin{abstract}
Collective electron dynamics in plasmonic nanosystems can unfold on timescales in the attosecond regime and the direct measurements of plasmonic near-field
oscillations is highly desirable. We report on numerical studies on the
application of attosecond nanoplasmonic streaking spectroscopy to the
measurement of collective electron dynamics in isolated Au nanospheres. The
plasmonic field oscillations are induced by a few-cycle NIR driving field and
are mapped by the energy of photoemitted electrons using a synchronized,
time-delayed attosecond XUV pulse.
By a detailed analysis of the amplitudes and phase shifts, we identify the different regimes of nanoplasmonic streaking and study the dependence on particle size, XUV photoelectron energy and emission position. The simulations indicate that the near-fields around the
nanoparticles can be spatio-temporally reconstructed and may give detailed insight into the build-up and decay of collective electron motion.
\end{abstract}

\maketitle
Nanoplasmonics has rapidly evolved and numerous techniques have been developed
to study the effects of plasmonic field enhancement \cite{Vogelsang10}, but so
far the direct, time-resolved measurement of plasmonic near-fields has not been
achieved. The fastest dynamics in plasmonic nanosystems can take place on
timescales down to 100 attoseconds as determined from the inverse bandwidth of
plasmonic resonance spectra. Attosecond metrology has provided valuable
tools for measurements of ultrafast electron dynamics in atoms \cite{Drescher02,
Uiberacker07, Schultze10, Goulielmakis10}, molecules \cite{Sansone10} and
surfaces \cite{Cavalieri07}.
One of the most successful techniques is attosecond streaking spectroscopy
\cite{Itatani02, Kienberger04}, employing photoemission of electrons by an
attosecond XUV pulse synchronized to a strong optical field.
Recording the electron kinetic energy spectra (by e.g. time-of-flight (TOF) spectroscopy) as a
function of the delay between the two pulses allows for the reconstruction of
the laser fields and the electron emission dynamics.
The technique is consequently a promising candidate for the real-time observation of collective electron motion in nanosystems.

One of the key aspects in traditional attosecond streaking on e.g. gas phase
atomic targets is the spatial homogeneity of the driving laser field.
In contrast, noble metal nanoparticles exhibit strongly enhanced, but
highly localized optical fields, such that the assumption of spatial homogeneity is no
longer valid.
When applying attosecond streaking spectroscopy to nanoparticles, the spatial decay of the near-field into free space will govern the streaking process. Nanoplasmonic streaking was originally proposed for the instantaneous electric field probing regime \cite{Stockman07}.
A study on integrated streaking spectroscopy on nanostructured antennas showed, that a reconstruction of the relatively homogeneous field in the antenna gap is possible if photoemission is limited to this region \cite{Skolapova2011}. Here the electron acceleration mostly takes place in the ponderomotive regime.

We extend this previous work to spherical Au nanoparticles of different sizes and exploring the transition from the instantaneous to the ponderomotive streaking regime. By taking into account the full spatial and temporal dependence of the plasmonic near-field, we analyze the streaking processes as a function of the nanoparticle size and laser parameters in detail. Spherical particles allow the use of Mie theory and were chosen for computational efficiency. The conclusions drawn in our study are, however, not limited to spheres, as the spatial decay of the near-field for a variety of systems shows similar, exponential behavior \cite{Jain2010}.
\begin{figure}[htp]
\centering
\includegraphics[scale=0.39,keepaspectratio=true]{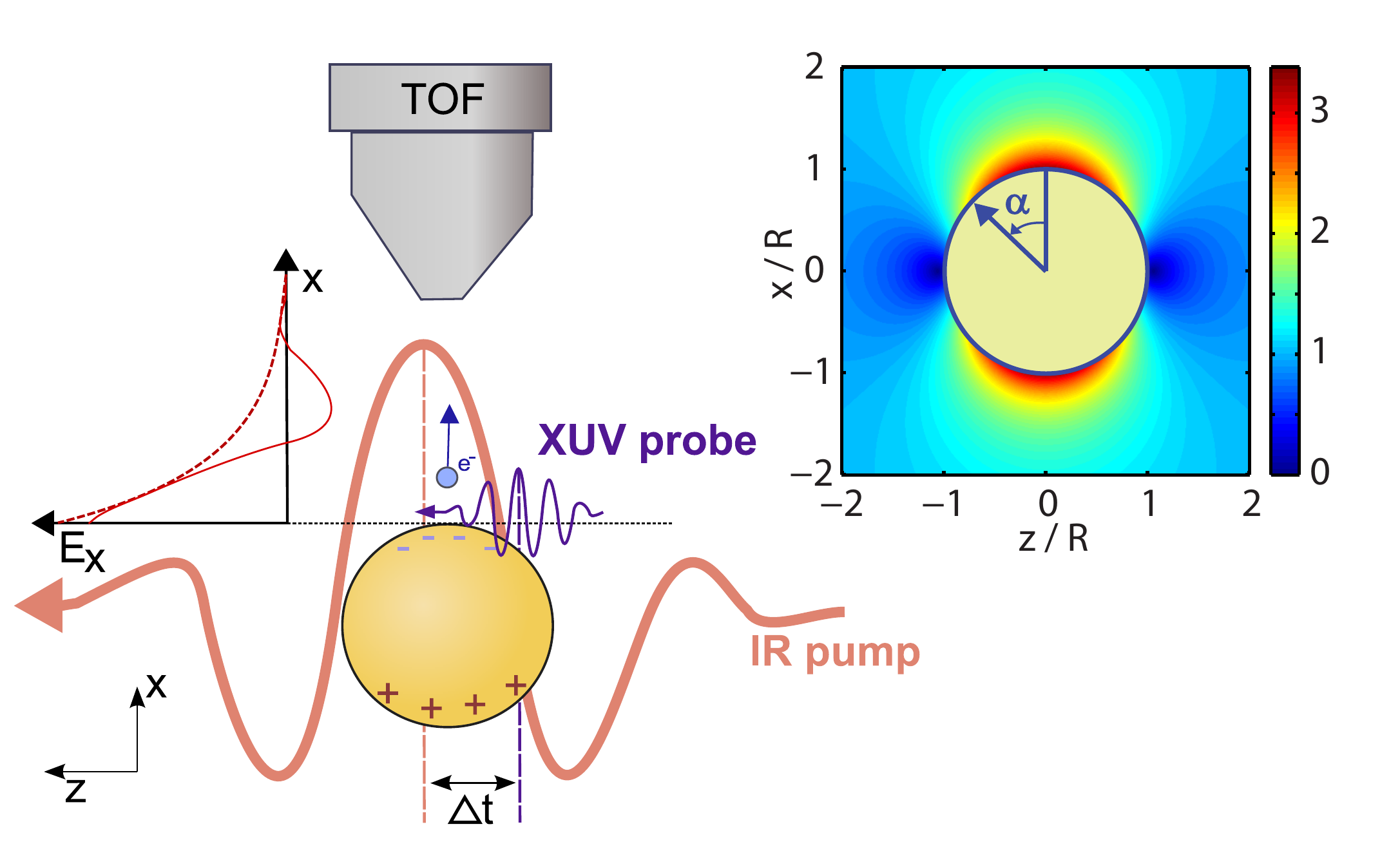}

\caption{(color online) Scheme for attosecond nanoplasmonic streaking of Au
nanospheres employing few-cycle NIR pump pulses, attosecond XUV probe pulses and
the measurement of the kinetic energies of photoemitted electrons using a
time-of-flight (TOF) spectrometer. The inset shows the field
enhancement as calculated by Mie theory for $y = 0$.
Due to symmetry of the plasmonic field, the point of electron emission is
uniquely defined by the angle $\alpha$. }
\label{fig:fig1}
\end{figure}
Employing isolated nanoparticles omits the requirement for spatially resolved experiments such as photoelectron emission microscopy, which is experimentally challenging in conjunction with attosecond light sources \cite{Mikkelsen2009,Chew2011}. Such isolated nanoparticles of nearly equal size and shape may be prepared by chemical synthesis in large quantities \cite{Lu09}. The nanoparticles can be supplied in sufficient density to the laser focus by employing either aerodynamic lenses \cite{Wang06,Zherebtsov11}, optical trapping \cite{Meinen10} or laser ablation \cite{Noel07}. From the experimental point of view, there are two major advantages of such an approach. First, the contrast in the spectra is expected to be higher, as this approach does not include a substrate. Furthermore, as the particles are replaced for each laser shot, the damage threshold of the Au nanospheres \cite{Plech2006} does not play a role and the non-linear plasmonic regime can be studied as well.

The scheme for attosecond nanoplasmonic streaking of fields near a metal
nanosphere is depicted in figure \ref{fig:fig1}. A few-cycle NIR laser pulse
excites the plasmonic oscillations and a time-delayed attosecond XUV laser pulse
(with a photon energy on the order of $\unit[100]{eV}$) photoemits electrons
from the nanosphere. The photoelectrons under consideration here are emitted
from the Fermi edge of the conduction band. Furthermore they are assumed to not have
scattered with other electrons or the lattice, which is justified if emission
takes place close to the surface as the electron's mean free path is on the
order of a few monolayers \cite{Tanuma91}. Accordingly the initial electron
kinetic energy is given by $E_{kin} = h\nu_{xuv} - W_f$ for the XUV frequency
$\nu_{xuv}$ and the work function of Au of $W_f = \unit[5.1]{eV}$.
While traveling to the detector, the electrons are accelerated by the local
electric field near the metal nanoparticle resulting in a kinetic energy shift
as a function of delay time that is measured with a TOF spectrometer. In our
simulations the final drift velocity of a photoelectron (emitted at time $t_e$ with an initial velocity $\vec{v}_0$)
was obtained by integrating the electron's classical equation of motion
\begin{equation}
\vec{v}_f(t_e) = \vec{v}_0 - \int_{t_e}^{\infty}dt \frac{\vec{E}(\vec{r},t) e}{m},
\label{eq_motion}
\end{equation}
where $e$ is the elementary charge and $m$ is the electron mass.
The integration was performed with an explicit Runge-Kutta algorithm (4th order)
using a time step of 10 as. The plasmonic response and the corresponding
near-field
$\vec{E}(\vec{r},t)$ is calculated using Mie theory \cite{Mie1908, LeRu2009}.
Using this approach, the enhancement and phase shift of the electric field
outside the sphere can be calculated by a normal mode expansion of the sphere's
electromagnetic response. The incident, linearly polarized laser field is
\begin{equation}
\vec{E}(t)=E_0 \mathrm{exp}\left(\frac{-t^2}{\tau^2}\right)\mathrm{cos}\left(\omega t
+ \phi\right)\hat{e}_x,
\label{gauss}
\end{equation}
with $\tau = \unit[2.1]{fs}$, corresponding to a full-width at half-maximum of 5
fs, an angular frequency $\omega$ corresponding to a center wavelength of
$\lambda$ = 720 nm, and a carrier-envelope phase $\phi = \pi$. The Coulomb field of the charged nanoparticle after electron emission is orders of magnitude smaller than the plasmonic near-field and thus neglected.

\begin{figure}[htp]
\centering
\includegraphics[scale=0.20,keepaspectratio=true]{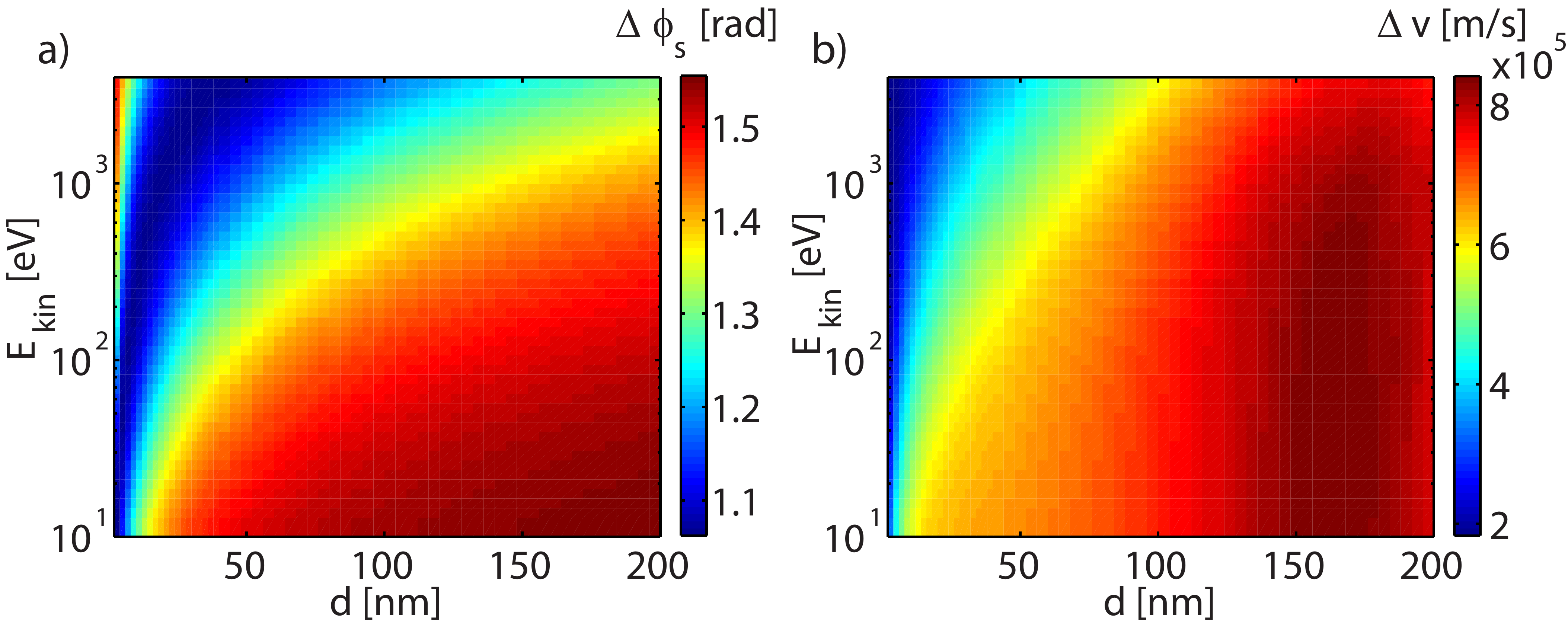}
\caption{(color online) a) Phase shift $\Delta \phi_{s}$ and b) maximum velocity
shift as a function of electron energy $E_{kin}$ and Au sphere diameter $d$ obtained for a
laser intensity of $\unit[10^{12}]{W/cm^2}$ and $\alpha = 0^\circ$. }
\label{fig:fig2}
\end{figure}
Using the appropriate wavelength dependent dielectric function for Au fitted to experimental data,
Mie theory predicts the resonance of Au spheres to be well below 600 nm
for diameters up to 100 nm (see Suppl. Inf.). In turn, for these
diameters the system is driven off resonance with the implication that the field
enhancement as well as the phase of the enhanced near-field do not vary
significantly over the entire NIR laser spectrum. Under this condition, the
calculations can be simplified and the entire laser spectrum treated with a
single set of Mie coefficients for the center wavelength.
The phase shift $\Delta \phi_{p}$ of the near-field with respect to the driving
field was taken into account in the calculations, but plays only a minor role
for small Au spheres. The field enhancement factor $\sigma = E_p / E_0$ for Au
spheres is predicted to be between 3.4 and 4 for spheres with diameters $d$
between $\unit[10]{nm}$ and $\unit[100]{nm}$, respectively. Throughout this
study we use parameters for which the maximum velocity shift of a photoelectron
is small compared to its initial velocity, so that recollision and rescattering
do not have to be taken into account. Electrons resulting from multi-photon emission and acceleration in the NIR field are
neglected since for the chosen NIR intensity they are energetically well separated from the XUV emitted electrons \cite{Dombi2009}.

First, we limit the analysis to electrons liberated from  $\alpha = 0^\circ$
(see fig. \ref{fig:fig1}) and traveling in x-direction corresponding to the
emission along the polarization axis towards the TOF spectrometer.
Significant insight can be gained from the phase shift between the streaking
curve and the plasmonic near-field $\Delta \phi_{s}$. The phase shift serves as
a measure how the streaking spectrum is generated and is in principle dependent
on all experimental parameters (namely sphere diameter, electron energy,
enhancement factor, duration of the near-field). Figure \ref{fig:fig2} a) shows $\Delta
\phi_{s}$ as a function of the sphere diameter $d$ and electron energy $E_{kin}$ for
$\tau = \unit[2.1]{fs}$ and an intensity of $\unit[10^{12}]{W/cm^2}$ of the NIR
field. The size dependence of $\Delta \phi_{s}$ originates from the spatial
extension of the near-fields into the vacuum, which is decaying with
approximately $1/r^3$, with $r$ being the radial distance from the origin. For
moderate electron energies and large spheres, the phase shift is approaching
$\pi / 2$.

In this regime the electron velocity is small
with respect to the field decay length, resulting in a ponderomotive
acceleration in the plasmonic field. Decreasing the particle size at a fixed
initial kinetic energy leads to a decrease of $\Delta \phi_{s}$. Now the regime
of instantaneous near-field probing is approached, as the electrons leave the
near-field faster and the acceleration is thus also confined in time. Going to
even smaller particles, $\Delta \phi_{s}$ eventually increases again, because
here the net acceleration by the near-field is smaller than the ponderomotive
acceleration in the laser field.
The minimum achievable phase shift is mainly determined by the enhancement
factor $\sigma$, which defines the relative strength of the two contributions.
The maximum velocity shift $\Delta v_{max}$ of the streaking curve is shown in
Figure \ref{fig:fig2} b). For the moderate near-field intensities used here, the
velocity shift increases linearly with the electric near-field amplitude.
Consequently, $\Delta v_{max}$ generally follows the field enhancement factor
peaking for a sphere diameter of about 170 nm. For small sizes and/or large
electron energies, $\Delta v_{max}$ approaches the value expected from streaking
in the homogenous laser field of $\unit[2\times 10^{-5}]{m/s}$.

For the spherical Au particles under consideration with relatively low
enhancement factors, we can conclude that a direct probing of the electric
field, as was discussed in Stockman \textit{et al.} \cite{Stockman07} is not
achievable. On the other hand, the enhancement factor is large enough to give
sufficient contrast in streaking spectrograms especially in the ponderomotive
regime.

\begin{figure}[htp]
\centering
\includegraphics[scale=0.33,keepaspectratio=true]{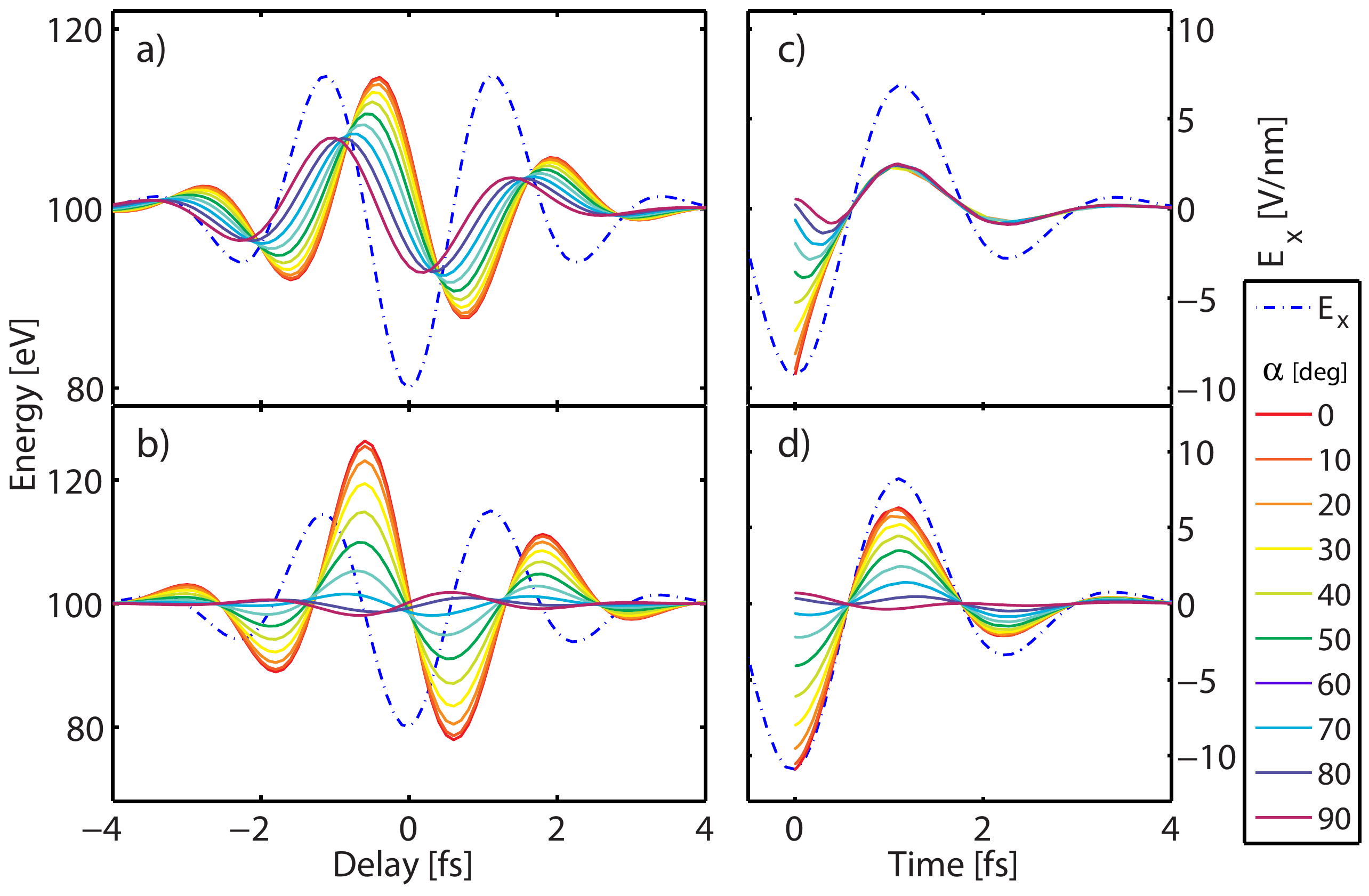}
\caption{(color online) Streaking traces for emission positions $\alpha =
0-90^\circ$ for 10 nm (a) and 100 nm spheres (b) and the respective electric
fields acting on the
photoelectrons (emitted at $t_e = \unit[0]{fs}$) as a function of time after
photoemission, (c) and (d). The blue dashed lines show the plasmonic near field.}
\label{fig:fig3}
\end{figure}

Another important aspect arising from the spatial inhomogeneity of the plasmonic
fields is the dependence of nanoplasmonic streaking on
initial conditions such as emission position and launch angle.
This is especially important, because XUV radiation can liberate electrons at
any position on the sphere. For the following discussion the XUV photon energy
is kept fixed at $h \nu_{xuv} = \unit[105]{eV}$, which can routinely be produced
via high-harmonic generation in Ne \cite{Goulielmakis08}. Figures
\ref{fig:fig3}a) and b) show the influence of the emission position $\alpha$
on the streaking traces for 10 nm and 100 nm Au spheres, respectively.
In general, for larger values of $\alpha$ (representing emission from a ring)
the phase shift $\Delta \phi_{s}$ increases while the streaking amplitude
decreases. This can be explained by the field distribution around the nanosphere
(see figure \ref{fig:fig1}).
Note that the particle sizes considered here are much smaller than the
wavelength, such that the dipolar mode
is dominating and all spheres show qualitatively similar field distributions.

Neglecting additional acceleration, the 100 eV photoelectrons travel about 6 nm
per femtosecond. Electrons emitted on-axis ($\alpha = 0^\circ$) will show the
largest energy shift and smallest phase shift $\Delta \phi_{s}$, as they are
accelerated by the strongest near-field. Photoelectrons emitted at the sides of
the sphere ($\alpha \approx 90^\circ$) initially experience weak or even
opposite electric fields (with respect to
$\alpha = 0^\circ$) after liberation.
The electric fields acting on electrons that have left the sphere from different
emission positions and travel to the detector will eventually converge.
Figure  \ref{fig:fig3}c) and d) show the effective,
time-dependent electric fields acting on electrons for 10 nm and 100 nm Au
spheres, respectively. For the 10 nm sphere the fields of all emission positions
have become equal after less than one half-cycle of the plasmonic field
oscillation.
For larger spheres this situation changes significantly. The electron travel
time through the near-field approximately scales with the sphere radius.
Electrons emitted from the side of e.g. a 100 nm sphere do not reach the regions
of strong field enhancement before the near-field oscillations vanish (owing to
the short pulse duration discussed here). Accordingly, these electrons do not
show noticeable net acceleration.

The complex dependence of the electron energy shift on the initial emission
conditions will determine nanoplasmonic streaking spectrograms recorded under
the described geometry. A distortion of the streaking spectrograms especially
for larger sizes is expected from the analysis above. To study these effects in
more detail, a large number of electron trajectories with randomized initial
conditions were integrated for
each sphere size. The initial conditions accounted for the probabilities of the
emission position, launch angle (within a 44$^\circ$ cone, corresponding to the
acceptance angle of the TOF spectrometer), kinetic energy within the XUV
spectral bandwidth (5~eV) and XUV pulse duration (200~as).
Transmission of XUV through the nanosphere and emission from the back side was
included in the calculations (see Suppl. Inf.).
The trajectory calculations were performed for the few-cycle NIR pulses
described above ($\tau = \unit[2.1]{fs}$, $I = \unit[10^{12}]{W/cm^2}, \phi =
\pi$).

\begin{figure}[htp]
\centering
\includegraphics[scale=0.24,keepaspectratio=true]{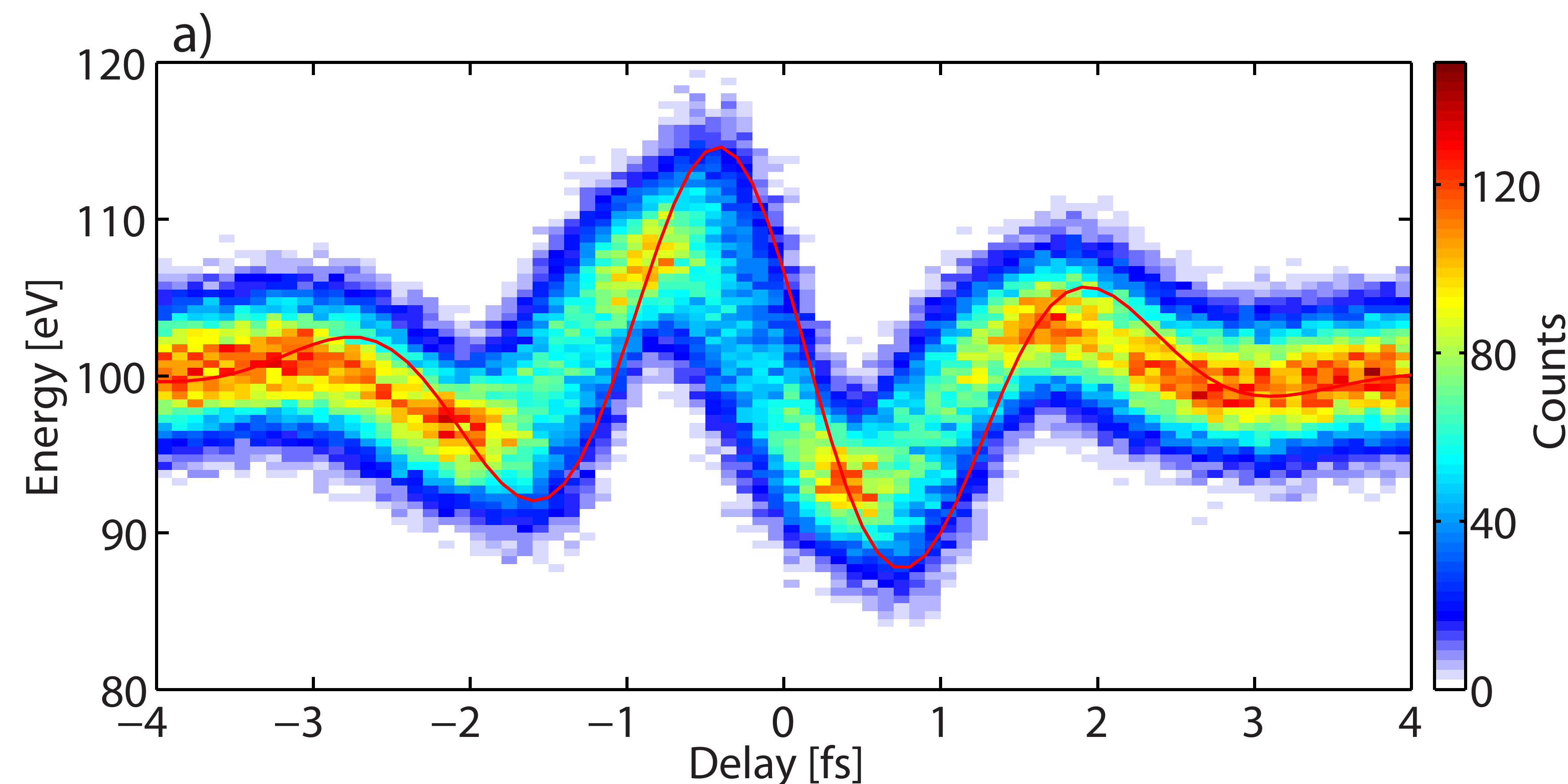}
\includegraphics[scale=0.24,keepaspectratio=true]{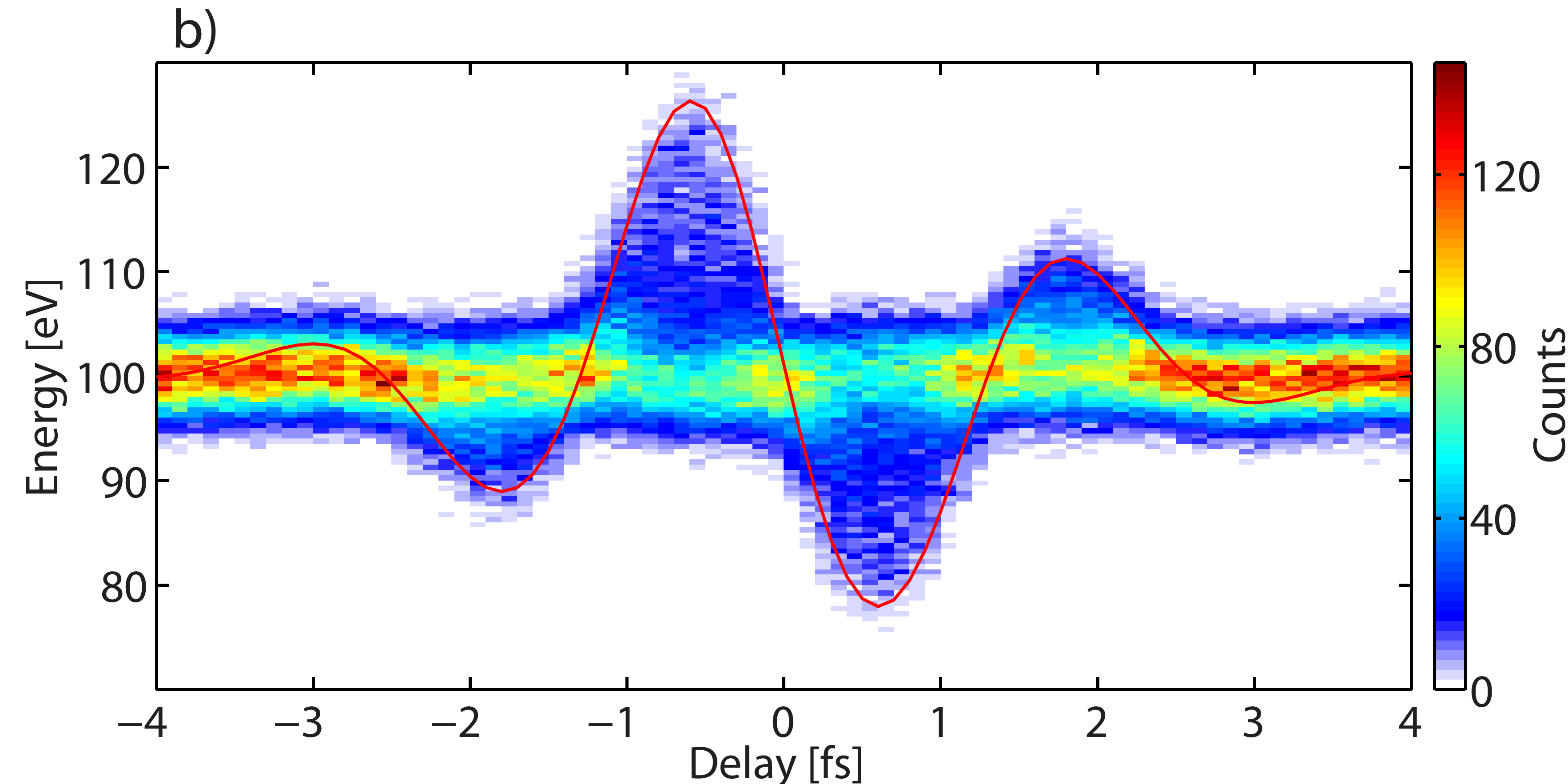}
\caption{(color online) Simulated streaking spectrograms for $\unit[10]{nm}$ (a)
and $\unit[100]{nm}$ (b) Au
spheres. The red line shows the contribution of the $\alpha = 0^\circ$
trajectories.}
\label{fig:fig4}
\end{figure}

Figures \ref{fig:fig4}a) and b) show streaking spectrograms composed of
1.5$\times$10$^5$ electron trajectories for $\unit[10]{nm}$ and $\unit[100]{nm}$
spheres, respectively. The two streaking spectrograms are vastly different. The
amplitude and phase shifts for different emission positions described above are
responsible for this behavior. The typical streaking pattern is almost
completely washed out in \ref{fig:fig4}b) ($\unit[100]{nm}$), while it can still
be observed in \ref{fig:fig4}a) ($\unit[10]{nm}$). Electrons emitted from
a $\unit[100]{nm}$ sphere at angles $\alpha$ larger than $60^\circ$ are hardly
accelerated. Since the statistical weight of each contribution to the
spectrogram is approximately proportional to the respective area on the sphere,
around 50\% of the photoemitted electrons exhibit minimal energy shifts.

A streaking spectrogram recorded for a rare gas (e.g.
Ne) \cite{Kienberger04} under the same experimental conditions can serve as a
reference permitting the precise reconstruction of the driving NIR field and XUV
pulse parameters \cite{Goulielmakis08}.
Such a reconstruction is also desirable for the temporal and spatial properties
of the plasmonic field. As the equation of motion of the electrons in its
general form (eq. \ref{eq_motion}) is not analytically integrable, numerical
simulations as presented in Fig. \ref{fig:fig4} have to be performed for the
retrieval of the spatio-temporal near-fields of the nanospheres. The
reconstruction could employ a feedback algorithm, which modifies the simulation
parameters until best agreement with the measured spectrogram has been achieved.

Under the assumption of a constant average electron velocity and approximating
the spatial near-field decay by an exponential function, an analytical
integration of the equation of motion for $\alpha = 0^\circ$ trajectories is
possible (see Suppl. Inf.). This allows for a computational less
expensive retrieval. A major difficulty, however, lies in extracting this
specific contribution from the streaking spectrogram. As can be seen in figure
\ref{fig:fig4}, the streaking waveform for $\alpha = 0^\circ$ (red line) can be
identified more easily for the 100 nm sphere and enables an analytic
reconstruction.

In order to test the experimental implementation of attosecond nanoplasmonic
streaking, we estimated the expected electron
count rates using table-top attosecond XUV light sources and typical
nanoparticle target densities provided by aerosol evaporation techniques \cite{Wang06}.
Although we have only considered single particles in our simulations, an ensemble of nanoparticles may be utilized in an experiment for a target that provides sufficient spatial separation between individual particles.
For an XUV pulse with $10^5$ photons, a target density of $\unit[10^{6}]{cm^{-3}}$ and $\unit[100]{nm}$ Au particles
we obtain a count rate of $\unit[1]{s^{-1}}$ using a 10\,kHz laser system (see Suppl. Inf.). The
count rate can be increased by one-three orders of magnitude by employing
optical trapping methods \cite{Meinen10} or laser ablation targets
\cite{Noel07}. Alternatively, an XUV light source with very high flux, such as
the one expected to emerge from the European Light Infrastructure (ELI)
\cite{Chambaret10}, could be utilized.

In conclusion, we analyzed and discussed the attosecond streaking of plasmonic
near-fields around isolated Au nanospheres of different diameters. In contrast
to conventional attosecond streaking, the electron acceleration and deceleration
has a strong spatial dependence and significant differences are identified in
our simulations for 10 nm and 100 nm particles. These differences are caused by
retardation effects between electron trajectories emitted from different
positions on the nanosphere. The spatio-temporal information that can be
retrieved from recorded streaking spectrograms could greatly deepen our
understanding of how plasmons form and how they decay. In contrast to studies on
nanostructured surfaces, where the field strength must remain below the damage
threshold of the material, a beam of isolated nanoparticles would allow studies
in the non-linear regime. Furthermore, the approach can be extended to other
material groups and be used to e.g. study – in real-time - ultrafast
field-induced phase transitions, such as the metallization of semiconductors and
dielectrics \cite{Durach10,Durach11}.

\begin{acknowledgments}
We acknowledge discussions of this work with T. Fennel, M. Stockman, and V.
Yakovlev. We are grateful for support by the DFG via SPP1391,
the Emmy-Noether program and the Cluster of Excellence: Munich Center for
Advanced Photonics.
\end{acknowledgments}

\bibliographystyle{apsrev}
\bibliography{complete}

\begin{thebibliography}{28}
\expandafter\ifx\csname natexlab\endcsname\relax\def\natexlab#1{#1}\fi
\expandafter\ifx\csname bibnamefont\endcsname\relax
  \def\bibnamefont#1{#1}\fi
\expandafter\ifx\csname bibfnamefont\endcsname\relax
  \def\bibfnamefont#1{#1}\fi
\expandafter\ifx\csname citenamefont\endcsname\relax
  \def\citenamefont#1{#1}\fi
\expandafter\ifx\csname url\endcsname\relax
  \def\url#1{\texttt{#1}}\fi
\expandafter\ifx\csname urlprefix\endcsname\relax\def\urlprefix{URL }\fi
\providecommand{\bibinfo}[2]{#2}
\providecommand{\eprint}[2][]{\url{#2}}

\bibitem[{\citenamefont{Vogelsang and Dimitriev}(2010)}]{Vogelsang10}
\bibinfo{author}{\bibfnamefont{R.}~\bibnamefont{Vogelsang}} \bibnamefont{and}
  \bibinfo{author}{\bibfnamefont{A.}~\bibnamefont{Dimitriev}},
  \bibinfo{journal}{Analyst} \textbf{\bibinfo{volume}{135}},
  \bibinfo{pages}{1175} (\bibinfo{year}{2010}).

\bibitem[{\citenamefont{Drescher et~al.}(2002)}]{Drescher02}
\bibinfo{author}{\bibfnamefont{M.}~\bibnamefont{Drescher}}
  \bibnamefont{et~al.}, \bibinfo{journal}{Nature}
  \textbf{\bibinfo{volume}{419}}, \bibinfo{pages}{803} (\bibinfo{year}{2002}).

\bibitem[{\citenamefont{Uiberacker et~al.}(2007)}]{Uiberacker07}
\bibinfo{author}{\bibfnamefont{M.}~\bibnamefont{Uiberacker}}
  \bibnamefont{et~al.}, \bibinfo{journal}{Nature}
  \textbf{\bibinfo{volume}{446}}, \bibinfo{pages}{627} (\bibinfo{year}{2007}).

\bibitem[{\citenamefont{Schultze et~al.}(2010)}]{Schultze10}
\bibinfo{author}{\bibfnamefont{M.}~\bibnamefont{Schultze}}
  \bibnamefont{et~al.}, \bibinfo{journal}{Science}
  \textbf{\bibinfo{volume}{328}}, \bibinfo{pages}{1658} (\bibinfo{year}{2010}).

\bibitem[{\citenamefont{Goulielmakis et~al.}(2010)}]{Goulielmakis10}
\bibinfo{author}{\bibfnamefont{E.}~\bibnamefont{Goulielmakis}}
  \bibnamefont{et~al.}, \bibinfo{journal}{Nature}
  \textbf{\bibinfo{volume}{466}}, \bibinfo{pages}{739} (\bibinfo{year}{2010}).

\bibitem[{\citenamefont{Sansone et~al.}(2010)}]{Sansone10}
\bibinfo{author}{\bibfnamefont{G.}~\bibnamefont{Sansone}} \bibnamefont{et~al.},
  \bibinfo{journal}{Nature} \textbf{\bibinfo{volume}{465}},
  \bibinfo{pages}{763} (\bibinfo{year}{2010}).

\bibitem[{\citenamefont{Cavalieri et~al.}(2007)}]{Cavalieri07}
\bibinfo{author}{\bibfnamefont{A.}~\bibnamefont{Cavalieri}}
  \bibnamefont{et~al.}, \bibinfo{journal}{Nature}
  \textbf{\bibinfo{volume}{449}}, \bibinfo{pages}{1029} (\bibinfo{year}{2007}).

\bibitem[{\citenamefont{Itatani et~al.}(2002)}]{Itatani02}
\bibinfo{author}{\bibfnamefont{J.}~\bibnamefont{Itatani}} \bibnamefont{et~al.},
  \bibinfo{journal}{Phys. Rev. Lett.} \textbf{\bibinfo{volume}{88}},
  \bibinfo{pages}{173903} (\bibinfo{year}{2002}).

\bibitem[{\citenamefont{Kienberger et~al.}(2004)}]{Kienberger04}
\bibinfo{author}{\bibfnamefont{R.}~\bibnamefont{Kienberger}}
  \bibnamefont{et~al.}, \bibinfo{journal}{Nature}
  \textbf{\bibinfo{volume}{427}}, \bibinfo{pages}{817} (\bibinfo{year}{2004}).

\bibitem[{\citenamefont{Stockman et~al.}(2007)}]{Stockman07}
\bibinfo{author}{\bibfnamefont{M.~I.} \bibnamefont{Stockman}}
  \bibnamefont{et~al.}, \bibinfo{journal}{Nature Phot.}
  \textbf{\bibinfo{volume}{1}}, \bibinfo{pages}{539} (\bibinfo{year}{2007}).

\bibitem[{\citenamefont{Skopalova et~al.}(2011)}]{Skolapova2011}
\bibinfo{author}{\bibfnamefont{E.}~\bibnamefont{Skopalova}}
  \bibnamefont{et~al.}, \bibinfo{journal}{New J. Phys.}
  \textbf{\bibinfo{volume}{13}}, \bibinfo{pages}{083003}
  (\bibinfo{year}{2011}).

\bibitem[{\citenamefont{Jain and {El-Sayed}}(2010)}]{Jain2010}
\bibinfo{author}{\bibfnamefont{P.~K.} \bibnamefont{Jain}} \bibnamefont{and}
  \bibinfo{author}{\bibfnamefont{M.~A.} \bibnamefont{{El-Sayed}}},
  \bibinfo{journal}{Chem. Phys. Lett.} \textbf{\bibinfo{volume}{487}},
  \bibinfo{pages}{153} (\bibinfo{year}{2010}).

\bibitem[{\citenamefont{Mikkelsen et~al.}(2009)}]{Mikkelsen2009}
\bibinfo{author}{\bibfnamefont{A.}~\bibnamefont{Mikkelsen}}
  \bibnamefont{et~al.}, \bibinfo{journal}{Rev. Sci. Instr.}
  \textbf{\bibinfo{volume}{80}}, \bibinfo{pages}{123703}
  (\bibinfo{year}{2009}).

\bibitem[{\citenamefont{Chew et~al.}(2011)}]{Chew2011}
\bibinfo{author}{\bibfnamefont{S.}~\bibnamefont{Chew}} \bibnamefont{et~al.},
  \bibinfo{journal}{submitted}  (\bibinfo{year}{2011}).

\bibitem[{\citenamefont{Lu et~al.}(2009)}]{Lu09}
\bibinfo{author}{\bibfnamefont{X.}~\bibnamefont{Lu}} \bibnamefont{et~al.},
  \bibinfo{journal}{Annu. Rev. Phys. Chem.} \textbf{\bibinfo{volume}{60}},
  \bibinfo{pages}{167} (\bibinfo{year}{2009}).

\bibitem[{\citenamefont{Wang and McMurry}(2006)}]{Wang06}
\bibinfo{author}{\bibfnamefont{X.}~\bibnamefont{Wang}} \bibnamefont{and}
  \bibinfo{author}{\bibfnamefont{P.}~\bibnamefont{McMurry}},
  \bibinfo{journal}{Aerosol Sci. Technol.} \textbf{\bibinfo{volume}{40}},
  \bibinfo{pages}{320} (\bibinfo{year}{2006}).

\bibitem[{\citenamefont{Zherebtsov et~al.}(2011)}]{Zherebtsov11}
\bibinfo{author}{\bibfnamefont{S.}~\bibnamefont{Zherebtsov}}
  \bibnamefont{et~al.}, \bibinfo{journal}{Nature Phys.}
  \textbf{\bibinfo{volume}{7}}, \bibinfo{pages}{656} (\bibinfo{year}{2011}).

\bibitem[{\citenamefont{Meinen et~al.}(2010)}]{Meinen10}
\bibinfo{author}{\bibfnamefont{J.}~\bibnamefont{Meinen}} \bibnamefont{et~al.},
  \bibinfo{journal}{Rev. Sci. Instr.} \textbf{\bibinfo{volume}{81}},
  \bibinfo{pages}{085107} (\bibinfo{year}{2010}).

\bibitem[{\citenamefont{Noel et~al.}(2007)}]{Noel07}
\bibinfo{author}{\bibfnamefont{S.}~\bibnamefont{Noel}} \bibnamefont{et~al.},
  \bibinfo{journal}{Appl. Surf. Sci.} \textbf{\bibinfo{volume}{253}},
  \bibinfo{pages}{6310} (\bibinfo{year}{2007}).

\bibitem[{\citenamefont{Plech et~al.}(2006)}]{Plech2006}
\bibinfo{author}{\bibfnamefont{A.}~\bibnamefont{Plech}} \bibnamefont{et~al.},
  \bibinfo{journal}{Nature Phys.} \textbf{\bibinfo{volume}{2}},
  \bibinfo{pages}{44} (\bibinfo{year}{2006}).

\bibitem[{\citenamefont{Tanuma et~al.}(1991)}]{Tanuma91}
\bibinfo{author}{\bibfnamefont{S.}~\bibnamefont{Tanuma}} \bibnamefont{et~al.},
  \bibinfo{journal}{Surf. Interface Anal.} \textbf{\bibinfo{volume}{17}},
  \bibinfo{pages}{911} (\bibinfo{year}{1991}).

\bibitem[{\citenamefont{Mie}(1908)}]{Mie1908}
\bibinfo{author}{\bibfnamefont{G.}~\bibnamefont{Mie}}, \bibinfo{journal}{Ann.
  Phys.} \textbf{\bibinfo{volume}{25}}, \bibinfo{pages}{377}
  (\bibinfo{year}{1908}).

\bibitem[{\citenamefont{Ru and Etchegoin}(2009)}]{LeRu2009}
\bibinfo{author}{\bibfnamefont{E.~C.~L.} \bibnamefont{Ru}} \bibnamefont{and}
  \bibinfo{author}{\bibfnamefont{P.~G.} \bibnamefont{Etchegoin}},
  \emph{\bibinfo{title}{Principles of Surface-Enhanced Raman Spectroscopy and
  Related Plasmonic Effects}} (\bibinfo{publisher}{Elsevier},
  \bibinfo{year}{2009}).

\bibitem[{\citenamefont{Dombi}(2009)}]{Dombi2009}
\bibinfo{author}{\bibfnamefont{P.}~\bibnamefont{Dombi}}, \bibinfo{journal}{Adv.
  Imaging and Electron Phys.} \textbf{\bibinfo{volume}{158}},
  \bibinfo{pages}{1} (\bibinfo{year}{2009}).

\bibitem[{\citenamefont{Goulielmakis et~al.}(2008)}]{Goulielmakis08}
\bibinfo{author}{\bibfnamefont{E.}~\bibnamefont{Goulielmakis}}
  \bibnamefont{et~al.}, \bibinfo{journal}{Science}
  \textbf{\bibinfo{volume}{320}}, \bibinfo{pages}{1614} (\bibinfo{year}{2008}).

\bibitem[{\citenamefont{Chambaret et~al.}(2010)}]{Chambaret10}
\bibinfo{author}{\bibfnamefont{J.-P.} \bibnamefont{Chambaret}}
  \bibnamefont{et~al.}, \bibinfo{journal}{Proc. SPIE}
  \textbf{\bibinfo{volume}{7721}}, \bibinfo{pages}{1D} (\bibinfo{year}{2010}).

\bibitem[{\citenamefont{Durach et~al.}(2010)}]{Durach10}
\bibinfo{author}{\bibfnamefont{M.}~\bibnamefont{Durach}} \bibnamefont{et~al.},
  \bibinfo{journal}{Phys. Rev. Lett.} \textbf{\bibinfo{volume}{105}},
  \bibinfo{pages}{086803} (\bibinfo{year}{2010}).

\bibitem[{\citenamefont{Durach et~al.}(2011)}]{Durach11}
\bibinfo{author}{\bibfnamefont{M.}~\bibnamefont{Durach}} \bibnamefont{et~al.},
  \bibinfo{journal}{Phys. Rev. Lett.} p. \bibinfo{pages}{in press}
  (\bibinfo{year}{2011}).

\end{thebibliography}

\end{document}